\documentclass[conference]{IEEEtran}
\IEEEoverridecommandlockouts
\usepackage{amsmath}
\usepackage{amsfonts}
\usepackage{amssymb}
\usepackage[ruled,vlined]{algorithm2e}
\usepackage{caption}
\usepackage{mathtools}  
\usepackage{amsthm}
\usepackage[]{algorithm2e}
\usepackage{booktabs}
\usepackage{subfigure}
\usepackage{tabulary}
\usepackage{footnote}
\usepackage[export]{adjustbox}
\usepackage{booktabs}
\usepackage[justification=centering]{caption}
\usepackage{comment}

\usepackage{cases}
\usepackage{graphicx}
\usepackage{cite}
\usepackage{multicol}
\usepackage{xcolor}
\usepackage{graphicx}

\title{Enhancing WiFi Multiple Access Performance with Federated Deep Reinforcement Learning}

\author{\text{Lyutianyang Zhang}$^1$\footnote{$^1$ \text{These authors contribute equally to this paper.}}, \text{Hao Yin}$^1$, \text{Zhanke Zhou}, \text{Sumit Roy}, \IEEEmembership{Fellow,~IEEE} \text{and Yaping Sun} 
\thanks{\textsuperscript{1}Both authors contributed equally to this work. Lyutianyang Zhang, Hao Yin, and Sumit Roy are with Department of Electrical \& Computer Engineering, University of Washington, Seattle, WA, USA (e-mail:\{lyutiz, haoyin, sroy\}@uw.edu). Zhanke Zhou is with School of Electronic Information and Communication, Huazhong University of Science and Technology, Wuhan, China (e-mail: zhankezhou@hust.edu.cn). Yaping Sun is with Cooperative Medianet Innovation Center, Shanghai Jiao Tong University, Shanghai 200240, China (e-mail: \{yapingsun\}@sjtu.edu.cn).}}%
\date{February 2020}

\begin{document}
 \pagenumbering{gobble}

\maketitle

\begin{abstract}
Carrier sensing multiple access/collision avoidance (CSMA/CA) is the backbone MAC protocol for IEEE 802.11 networks. However, tuning the binary exponential back-off (BEB) mechanism of CSMA/CA in user-dense scenarios so as to maximize aggregate throughput still remains a practically essential and challenging problem. In this paper, we propose a new and enhanced multiple access mechanism based on the application of deep reinforcement learning (DRL) and Federated learning (FL). A new Monte Carlo (MC) reward updating method for DRL training is proposed and the access history of each station is used to derive a DRL-based MAC protocol that improves the network throughput vis-a-vis the traditional distributed coordination function (DCF). Further, federated learning (FL) is applied to achieve fairness among users. The simulation results showcase that the proposed federated reinforcement multiple access (FRMA) performs better than basic DCF by 20\% and DCF with request-to-send/clear-to-send (RTS/CTS) by 5\% while guaranteeing the fairness in user-dense scenarios.

\end{abstract}
\section{Introduction} \label{introduction}

 The 802.11 WLAN base medium access control (MAC) - the so-called distributed coordination function (DCF) -  allows multiple mobile stations to share the common channel via a random distributed access time-slotted protocol that operates on the principle of Carrier Sense Multiple Access with Collision Avoidance (CSMA/CA). Increasing network throughput in a single WiFi cell/Basic service set identifier (BSSID) for an increasing number of users continues to be an important resource allocation problem.

As is well-known, all nodes in a CSMA/CA network are time-synchronized (hence enabling slotted operation) and employ a backoff mechanism while seeking channel access. Should two nodes simultaneously transmit, it leads to a (synchronous) MAC collision, whereupon the involved nodes employ a Binary Exponential Backoff (BEB) \cite{kwak2005performance} mechanism for mitigating subsequent collision events on channel access. 


There have been several studies optimizing 802.11 network throughput by tuning BEB. As a groundwork, in \cite{kwak2005performance,bianchi2000performance}, the performance of CSMA/CA with BEB was analyzed using the 1-D and 2-D Markov chains, respectively. The analytical results show that with the increasing number of stations, the aggregate network throughput decreases for base DCF and may be stabilized via the use of RTS/CTS. However, the search for alternatives and enhancements to BEB so as to further enhance network throughput while the number of users scale has continued \cite{li2009new,al2014enhanced}.
Furthermore, BEB is well-known to result in short-term unfairness whereby backed off nodes (due to repeated collisions) experience considerable delays for channel access. 

As for DCF with RTS/CTS, while it is usually helpful in improving and stabilizing aggregate throughput, the gains are very dependant on specific network topology and traffic conditions. In summary, there still exists a significant need for new robust, practical solutions for a distributed MAC protocol for 802.11 access that works well in a broad set of realistic operational scenarios.

In this work, we propose a new distributed MAC protocol for 802.11 networks, whereby all stations employ a neural network based on DRL principles\cite{watkins1992q,schmidhuber2015deep} and the AP adopts Federated Learning (FL) \cite{mcmahan2017federated,jiang2019improving}. Each station trains its parameters, i.e., the weights of neural network layers, by utilizing its access history including past channel states (Busy or Idle) and its corresponding action (Transmit or Wait) plus the feedback from Access Point (AP) corresponding to a certain transmit action of the station. Further, in order to achieve fairness of the distributed DRL-based MAC protocol, we incorporate federated learning (FL) principles.


In \cite{yu2019deep}, multiple access based on DRL is proposed. Although this work also aims to design a MAC protocol with higher throughput while guaranteeing the $\alpha$-fairness, it is based on Reinforcement Learning with centralized training and hence inapplicable to the WiFi network. On the contrary, we propose a {\em distributed DRL}  method in which each station has its personalized Q neural network (QNN). Further, we integrate FL to achieve faster convergence during training and proportional fairness.

The contributions of the paper are as follows:

\begin{itemize}
  \item We construct a framework for distributed DRL for the next generation of WiFi MAC protocol.
  \item We propose a new MC reward updating method for the training phase of DRL, and accelerate it using FL.
  \item The FRMA protocol improves throughput performance relative to DCF and also preserves access fairness with the same data rate.
\end{itemize}

The paper is organized as follows. The system model for each node as a DRL agent is introduced in Sec. II. The MC reward update method is then introduced for the fast and accurate training for DRL neural network. Then FL is introduced as the consensus mechanism for the AP to reach a global optimal based on all current trained DRL models. Sec. III presents a simulation-based performance evaluation of base DCF and DCF with RTS/CTS as the benchmark to enable comparisons with the performance of the proposed FRMA.  
\section{System Model}
\label{sec:sys}

\begin{figure}[t]
    \centering
    \includegraphics[width=.45\textwidth]{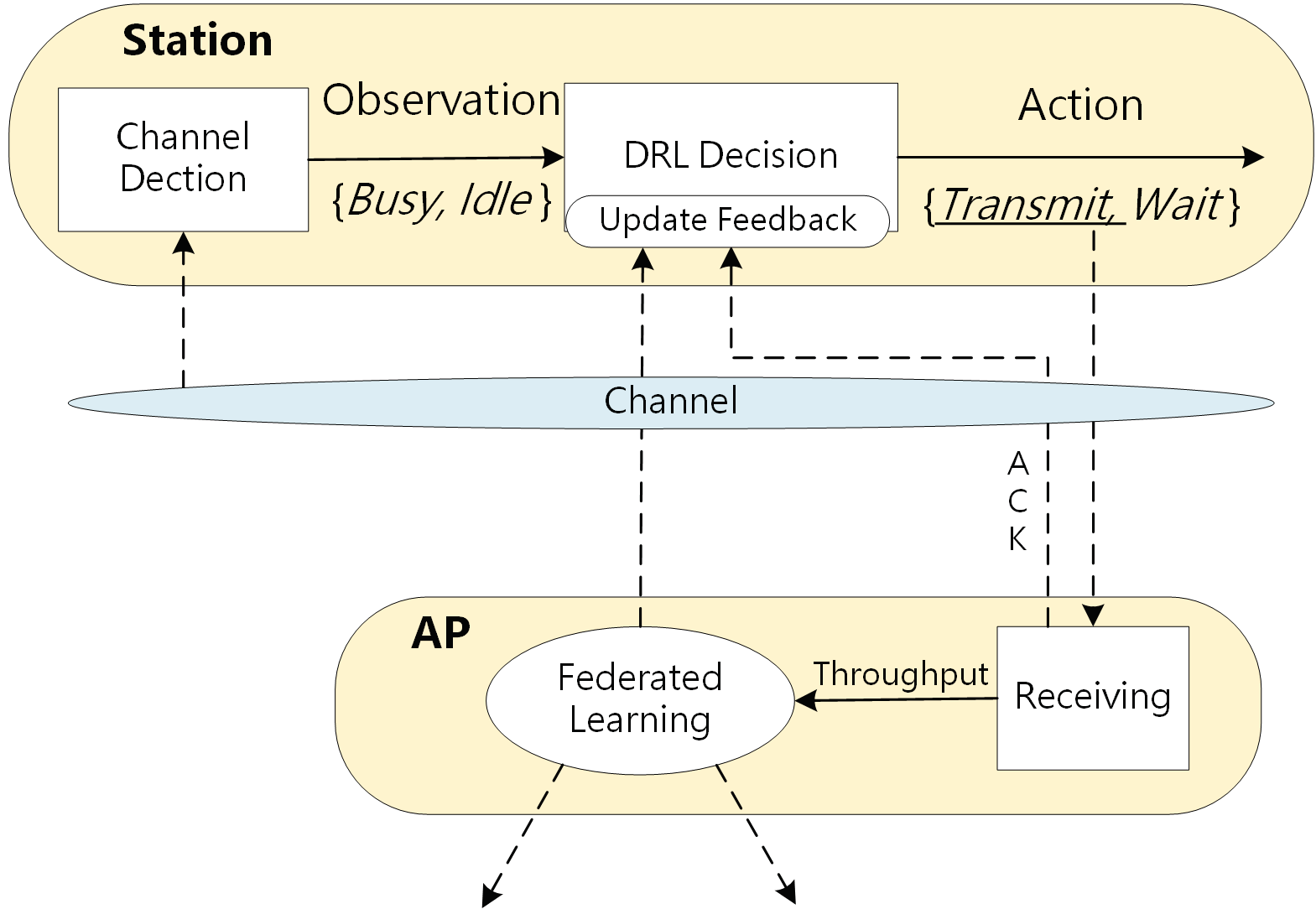}
    \caption{System operation in proposed MAC for stations and AP.}
    \label{fig:sim}
\end{figure}

The MAC layer elements in our proposed design for stations and AP are shown in Fig. \ref{fig:sim}. The MAC state description can be used to
construct a Markov decision process (MDP) formulation. Next, we introduce a Monte Carlo (MC) reward estimation method and standard Reinforcement Learning (RL) solution for each station. The AP implements Federated Learning (FL) for access fairness and fast convergence. In this for simplicity, we only consider uplink communication and that each station is always in saturation mode (always has data to be transmitted to the AP). 
\begin{algorithm}[t]
 \KwData{$s^{i}_{t}=\{c^{i}_{t-M+1},\dots,c^{i}_{t}\}$, $c^{i}_{t}=\{a^{i}_{t-1},o^{i}_{t}\}$, and $a_{t}^{i}=\underset{a_{t}^{i}}{\operatorname{argmax}}~q(s^{i}_{t},a_{t}^{i};\theta^{i})=\{Wait\}$}
 \KwResult{$r^{i}_{t+1}$}
 \eIf{$o^{i}_{t} == \{\text{Busy}\}$}
 {
 reward = 1\;
 }
 {
 reward = 0\; 
 }
 
 \caption{Reward estimation method 1 when QNN output is $\{Wait\}$ on Station $i$.}
\end{algorithm}

\begin{algorithm}[t]
 \KwData{$s^{i}_{t}=\{c^{i}_{t-M+1},\dots,c^{i}_{t}\}$, $c^{i}_{t}=\{a^{i}_{t-1},o^{i}_{t}\}$, $a_{t}^{i}=\underset{a_{t}^{i}}{\operatorname{argmax}}~q(s^{i}_{t},a_{t}^{i};\theta^{i})=\{Transmit\}$}
 \KwResult{$r^{i}_{t+1}$}
 
 $\mathcal{K}=\{\}$\;
 $l=t-M+1$\;
 \While{$l<=t$}{
  \eIf{$c^{i}_{l}==c^{i}_{t}$}{ $\mathcal{K}=\{\mathcal{K},c^{i}_{l}\}$\;}{$\mathcal{K}=\mathcal{K}$\;}
  $l=l+1$\;
 }
 $l=t-M+1$\;
 $reward=0$\;
 \While{$\mathcal{K}(l)$==True and $\mathcal{K}(l)$ exist}{
 $k=\mathcal{K}(l)$\;    
 \eIf{The feedback of $\{a^{i}_{k}\}=\{Transmit\}$, i.e., the feedback received at time slot $k^{\prime}$ is ACK.}{
 $reward=\eta \times reward +1$\; 
 }
 {
 $reward=\eta \times reward -1$\; 
 }
 $l=l+1$;
 }
 \caption{Monte Carlo (MC) reward estimation method 2 when QNN output is $\{Transmit\}$ on Station $i$. $\eta$ is the memory factor; the closer the same action happens to the current state, we allocate more weight to the reward of that state by forgetting the previous accumulative reward. Note that $k^{\prime}$ denotes the corresponding time slot when the transmission feedback information of the action $a^{i}_{k}$ is received by station $i$, if the result is ACK, we reward with $1$ and $-1$ if NACK.}
\end{algorithm}

\subsection{Markov Decision Process and Monte Carlo Reward Estimation Method}
Suppose there are $N$ stations, each of which seeks channel access to send data to AP, as shown in Fig. \ref{fig:sim}. The network is time-slotted (i.e, $t=0,1,\dots$) and at time slot $t$, station $i$ observes the current channel state,  $o^{i}_t \in \{\{Busy\},\{Idle\}\}$, and utilized the historic channel states and its own past actions, to decide the next action $a^{i}_t \in \mathcal{A}$,\\ where $\mathcal{A} = \{Transmit,Wait\}$. Although we give the general definition for channel state observation, in this paper, we assume that there is no hidden terminal problem, i.e., each station's observation $o^{i}_t$ is assumed to be the same as each other. The action-observation tuple at time slot $t+1$ for station $i$ is defined as $c^{i}_{t+1}=\{a^{i}_t,o^{i}_{t+1}\}$. Denote the environment history at time $t+1$ as $s^{i}_{t+1}=[c^{i}_{t-M+2},\dots,c^{i}_{t},c^{i}_{t+1}]$ which is a concatenation of channel states for the past $M$ slots, where $M$ denotes the experience replay memory length. Based on $s^{i}_{t}$, the station makes an action $a^{i}_{t}$ and the state transfers to $s^{i}_{t+1}$ with reward $r^{i}_{t+1} \in \mathcal{R}$. Note that the reward cannot be obtained immediately after the action since it is only decided after the feedback from the AP. For example, when $a^{i}_{t}=\{Transmit\}$, station $i$ observes the feedback (either timeout or acknowledgement) after a delay of $k$ slots, upon which the reward can be calculated. Therefore, once time out, we should punish the action $a^{i}_{t}$ ($r_{t+1}<0$) and reward ($r^{i}_{t+1}>0$) if ACK. The action can be punished/rewarded by looping through the historic information. Hence, the reward $r^{i}_{t+1}$ is decided by environment state $s^{i}_{t+1}$ and all previous states. We propose two reward estimation methods based on Monte Carlo method when $a^{i}_{t}=\{Transmit\}$ and $a^{i}_{t}=\{Wait\}$ in Algorithm 1 and 2 respectively. Algorithm 1 only requires the local information of the mobile station $i$ itself for reward estimation. Algorithm 2 requires the local information and the feedback information from AP for execution.

\subsection{Reinforcement Learning}
In this section, we introduce standard Q-learning and $\epsilon$-greedy policy as the foundation for the following Deep Reinforcement learning. As defined in the above section, each action $a^{i}_{t}$ transfers the current state of station i, $s^{i}_{t}$, to $s^{i}_{t+1}$ with reward $r^{i}_{t+1}$. The cumulative reward of station $i$ at time $t$ is then expressed as 
\begin{equation}
    R^{i}_{t}=\sum_{k=0}^{\infty} \gamma^{k} r^{i}_{t+k+1},
\end{equation}
where $\gamma \in \left(0,1\right]$ is a discounting factor.  Denote the policy function of station $i$ as $\pi(i)$, which identifies the action taken given a certain state. Now, we introduce the action-value function of station i as follows,
\begin{equation}
    Q^{\pi(i)}(s^{i},a^{i})=\mathbf{E}[R^{i}_t|s^{i}_t=s^{i},a_t=a^{i},\pi(i)].
\end{equation}

In order to maximize the cumulative reward, we need to find the optimal policy. The optimal action-value function, $Q^{\pi(i)}_{opt}(s^{i},a^{i})=\text{max}_{\pi(i)}Q^{\pi(i)}(s^{i},a^{i})$ obeys the Bellman optimality equation
\begin{equation}
\begin{aligned}
    &Q^{\pi(i)}_{opt}(s^{i},a^{i})\\&=\mathbf{E}_{s^{i}_{t+1}}[r^{i}_{t+1}+\gamma \max_{a^{i}_{t+1}} Q^{\pi(i)}_{opt}(s^{i}_{t+1},a^{i}_{t+1})|s^{i}_{t}=s^{s}, a^{i}_{t}=a^{i}],
\end{aligned}
\end{equation}
Q-learning is an iterative algorithm which estimates the action-value function $q(s^{i}_{t},a^{i}_t)$ via the following update 
\begin{equation}
    \begin{aligned}
        q(s^{i}_{t},a^{i}_t) \leftarrow & q(s^{i}_{t},a^{i}_t)\\ &+\beta \left[r^{i}_{t+1}+\gamma \max_{a^{i}_{t+1}}q(s^{i}_{t+1},a^{i}_{t+1})-q(s^{i}_{t},a^{i}_{t})\right],
    \end{aligned}
\end{equation}
where $\beta \in \left(0,1\right]$ denotes the learning rate. 

While each node updates $q(s^{i}_{t},a^{i}_{t})$, it also makes decisions based on $q(s^{i}_{t},a^{i}_{t})$. For the $\epsilon$-greedy policy, the optimal action is given by 
\begin{equation}
        a^{i}_{t+1}=
        \begin{cases}
         \text{argmax}_{a^{i}_{t+1}}q(s^{i}_{t+1},a^{i}_{t+1}), &P=1-\epsilon.\\
         \text{random action}, &P= \epsilon,
        \end{cases}
\label{eq:eps}
\end{equation}
where $\epsilon$ denotes the probability of choosing random action. Such a greedy Q-learning policy has been shown to be effective in avoiding saddle points by trying random new actions.

\subsection{Semi-gradient Algorithm in Deep Q Learning}
Given the fact that we have access to reward information $r^{i}_{t}$, the QNN can be trained by minimizing prediction errors of $q(s^{i}_{t},a^{i}_{t};\text{\boldmath$\theta$}^{i})$ at each station and time slot, where $\text{\boldmath$\theta$}^{i}$ denotes the weights-to-be-trained of QNN at station $i$. As introduced beforehand, action $a^{i}_{t}$ transfers $s^{i}_{t}$ to $s^{i}_{t+1}$ which gives the reward $r^{i}_{t+1}$. Thus, the above information, $(s^{i}_{t},a^{i}_{t},r^{i}_{t+1},s^{i}_{t+1})$, forms a single batch of training set for QNN. Next, we define the prediction loss function of QNN as 
\begin{equation}
    L(\text{\boldmath$\theta$}^{i})=(v(r^{i}_{t+1},s^{i}_{t+1})-q(s^{i}_{t},a^{i}_{t};\text{\boldmath$\theta$}^{i}))^2,
\end{equation}
where $q(s^{i}_{t},a^{i}_{t};\text{\boldmath$\theta$}^{i})$ is the output of QNN at time slot $t$ and the approximated value function is defined as 
\begin{equation}
    v(r^{i}_{t+1},s^{i}_{t+1})=r^{i}_{t+1}+\gamma \max_{a^{i}_{t+1}} q(s^{i}_{t+1},a^{i}_{t+1};\text{\boldmath$\theta$}),
    \label{eq:gam}
\end{equation}
where the second term $\gamma \max_{a^{i}_{t+1}} q(s^{i}_{t+1},a^{i}_{t+1};\text{\boldmath$\theta$})$ is obtained by searching the maximum output of QNN with respect to the selection of the action $a^{i}_{t+1}$ given $s^{i}_{t+1}$. Then, we can update using semi-gradient algorithm \cite{sutton2018reinforcement} as below,

\begin{equation}
\text{\boldmath$\theta$}^{i} \leftarrow \text{\boldmath$\theta$}^{i}+\rho\left[v(r^{i}_{t+1},s^{i}_{t+1})-q(s^{i}_{t+1},a^{i}_{t+1};\text{\boldmath$\theta$}^{i}) \right] \nabla q(s^{i}_{t},a^{i}_{t};\text{\boldmath$\theta$}^{i}),
\end{equation}
where $\nabla$ is the gradient with respect to $\text{\boldmath$\theta$}^{i}$.
\begin{algorithm}[h]
 \KwData{$\text{\boldmath$\theta^{i}$}$ }
 \KwResult{Broadcast $\text{\boldmath$\bar{\theta}$}$ to every station}
 \eIf{$mod(t,T)==T-1$}
 {
    $\text{\boldmath$\bar{\theta}$}=\sum_{i=1}^{N}\text{\boldmath$\theta^{i}$}$
 }
 {
Do nothing
 }
 \caption{FRMA at AP side. $mod(t,T)$ denotes $t$ modulo T, which means that the FedAvg mechanism is triggered every period of $T$.}
\end{algorithm}

\begin{figure*}[t]
    \centering
    \includegraphics[width=0.7\textwidth]{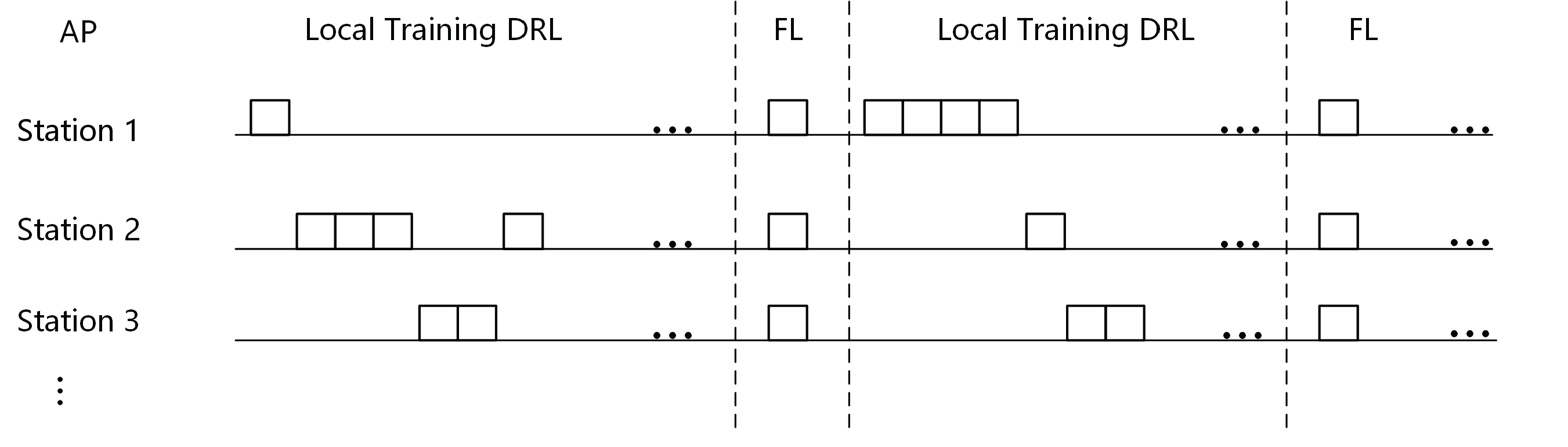}
    \caption{FRMA Protocol: each block represents a packet. In Local Training DRL, each station does QNN training by itself. However, during the FL timeframe, AP uses Algorithm 3 to obtain a global model of QNN and broadcast it to each station. It is noteworthy that the overhead of uploading QNN to AP and AP broadcasting global back to each station is almost negligible, which is discussed in the simulation section.}
    \label{fig:protocol}
\end{figure*}

\subsection{Federated Learning and FRMA}
Each station uses the described RL method for uplink random access starting with random initialization of QNN weight \text{\boldmath$\theta$}. A station is likely initialized with better QNN weights and a much more aggressive strategy than other stations. The consequence is that other stations will try to decrease their throughput to at least guarantee there is no collision, i.e., zero throughput for everyone. Therefore, if we employ only the aforementioned distributed RL network, we have no fairness at all. Hence, we add one more mechanism to this network, FL, which refers to learning a high-quality global model based on decentralized data storage. FL has been shown to be a fast convergent method in distributed networks \cite{mcmahan2017federated} for a large number of stations. We hereby propose our FL mechanism in Algorithm 3 at the AP side based on FedAvg \cite{mcmahan2016communication}. Based on the FL at the AP side and aforementioned distributed DRL, we propose a new MAC protocol, FRMA, shown in Fig. \ref{fig:protocol}.

\section{Performance Evaluation: Simulations}
\label{simulation}
In this section, we analyze the performance of FRMA, which searches for the policy maximizing the aggregated network throughput while guaranteeing the fairness among mobile stations. To compare with the most widely used CSMA/CA protocol, we also introduce and implement the simulation of DCF basic and RTS/CTS mechanism. We also use the Bianchi's theoretical model \cite{bianchi2000performance} to validate and calibrate the implementation. Based on this validated structure, we then implement our FRMA and performance comparison. 

\subsection{Performance review of IEEE 802.11 CSMA/CA}
\label{sec:multi}

In this section, we describe the analytical throughput estimation models of the CSMA/CA in IEEE 802.11 DCF network based on the well-known work \cite{bianchi2000performance} under the assumption of saturated networks, i.e., all nodes always have a packet to send. We consider both the packet transmission schemes employed by DCF, namely, the basic access and the request-to-send/clear-to-send (RTS/CTS) access mechanisms. The fraction of time the channel use
s to successfully transmit packets or the normalized average system throughput $S$ can be expressed as
\begin{equation}
\label{eq:fraction}
\begin{aligned}
    S&=\frac{\textbf{E}[\text { payload information transmitted in a slot time }]}{\textbf{E}[\text { length of a slot time }]}
    \\&=\frac{P_{s}P_{tr}\textbf{E}[P]}{ (1-P_{tr})\sigma + P_{tr}P_{s}T_{s} + P_{tr}(1-P_{s})T_{c}},\\
\end{aligned}
\end{equation}
where $T_s$ is the average time that the channel is busy due to a successful transmission, $T_c$ is the average time due to one packet collision, $P_{tr}$ is the probability of at least one station transmitting in a slot and $P_{s}$ is the probability of a successful transmission. Therefore, the actual throughput can be calculated as $S \times D$, where $D$ denotes the data rate.

The calculation of the transmitting time with DCF basic access and the RTS/CTS access mechanisms are shown in Eq \eqref{eq:t_bas} and \eqref{eq:t_rts}. For the basic DCF, the channel time for successful transmission $T_s = T_s^{bas}$ and collision $T_c = T_c^{bas}$, respectively. For the RTS/CTS Access mechanism, collision can occur only when RTS frames are transmitted, so we have $T_s = T_s^{rts}$ and $T_c = T_c^{rts}$.

\begin{equation}
\label{eq:t_bas}
\left\{\begin{array}{l}
T_{s}^{\mathrm{bas}}=H+\textbf{E}[P]+\mathrm{SIFS}+\delta+\mathrm{ACK}+\mathrm{DIFS}+\delta \\
T_{c}^{\mathrm{bas}}=H+\textbf{E}\left[P\right]+\mathrm{EIFS}+\delta
\end{array}\right.\end{equation}

\begin{equation}
\label{eq:t_rts}
\left\{\begin{aligned}
T_{s}^{\mathrm{rts}}=& \mathrm{RTS}+\mathrm{SIFS}+\delta+\mathrm{CTS}+\mathrm{SIFS}+\delta+H \\
&+\textbf{E}[P]+\mathrm{SIFS}+\delta+\mathrm{ACK}+\mathrm{DIFS}+\delta \\
T_{c}^{\mathrm{rts}}=& \mathrm{RTS}+\mathrm{EIFS}+\delta,
\end{aligned}\right.\end{equation}
where $H$ represents the MAC and PHY header time, and $\delta$ is the propagation delay equal to 0.1 $\mu$s, $\textbf{E}[P]$ is the expected value of the frame time duration, and ACK is the time for receiving ACK. If the channel is idle for a period of time equal to a distributed interframe space (DIFS), and the ACK is immediately transmitted at the end of the packet, after a period of time called short interframe space (SIFS). For the failure of the transmission, the EIFS = SIFS + ACK + $\delta$ is used instead of DIFS. For the RTS/CTS mechanism, the RTS and CTS repersents the request-to-send and clear-to-send time.  

Each station needs to wait for a random backoff time before transmitting it. The backoff is performed in discrete time units called slots, and the stations are synchronized on the slot boundaries. Using the 2-D Markov model developed in \cite{bianchi2000performance}, the probability that a station transmits in a time slot can be obtained as given by Eq \eqref{eq:mav}:
\vspace{-1mm}
\begin{equation}
    \tau = \frac{2}{W_0\left(\frac{(1-(2P_w)^{m+1})(1-P_w)+2^{m}\left(P_w^{m+1}-P_w^{R+1}\right)(1-2P_w)}{(1-2P_w)(1-P_w^{R+1})}\right) +1},
    \label{eq:mav}
\end{equation}
where $R$ is the number of the backoff stage, $W_{0}$ is the minimum contention window size, $m = log_{2}(\frac{CW_{max}}{CW_{min}})$ where $CW_{max}$ and $CW_{min}$ denotes the maximum and minimum length of the contention window size and respectively. Each frame collides with constant collision probability $p$ given as follows:
\vspace{-2.8mm}
\begin{equation}
 P_{w}=1-(1-\tau)^{n-1},
 \label{eq:colli}
\end{equation}
where n is the number of competing stations.

It is obvious that $\tau$ and $p$ can be obained by solving Eq. \eqref{eq:mav} and Eq. \eqref{eq:colli}. The $P_{tr}$ and $P_{s}$ for DCF baisc access and the RTS/CTS access mechanisms are expressed as follows: \eqref{eq:p_bas}.
\begin{equation}
    \begin{aligned}
& P_{tr}=1-(1-\tau)^{n} && \\
& P_{s}= \frac{n\tau(1-\tau)^{(n-1)}}{P_{tr}}. &&
\label{eq:p_bas}
\end{aligned}
\end{equation}

\subsection{Simulation Setup}
 
The simulation scenario is shown in Fig. \ref{fig:sim}. We utilize Python to implement the DRL and FL with PyTorch \cite{NEURIPS2019_9015} and estimate the aggregated throughput for different number of mobile stations. The simulations were implemented on a server with a CPU (Intel Core i9-9700k) and a GPU (NVIDIA GeForce GTX 2080Ti). The detailed simulation parameters and the hyper-parameters for the deep neural network we adopted for simulation are introduced in the following sections.

\subsubsection{Simulation Parameters}

We consider a single 802.11 network (BSSID) where all the stations are in saturation mode, i.e., they always compete for channel access. All stations are assumed to observe the same channel status in the same time slot and utilize the same transmission power and packet length. For base DCF and DCF with RTS/CTS, the system parameters are shown in table \ref{tb:sim_pra}. We conduct Monte-Carlo simulation with 100 independent trials with 200s simulation time and then the average of the results. 

\begin{table}[]
\caption{Simulation Parameters}
\label{tb:sim_pra}
\resizebox{.45\textwidth}{!}{%
\begin{tabular}{|c|c|c|c|}
\hline
\textbf{Parameters}         & \textbf{DCF Baisc} & \textbf{RTS/CTS} & \textbf{FRMA} \\ \hline
Slot time $\sigma$ ($\mu$s) & \multicolumn{3}{c|}{10}                                \\ \hline
SIFS ($\mu$s)               & \multicolumn{2}{c|}{16}               & NA             \\ \hline
DIFS ($\mu$s)               & \multicolumn{2}{c|}{34}               & NA             \\ \hline
RTS($\mu$s)                 & NA                 & 46               & NA             \\ \hline
CTS($\mu$s)                 & NA                 & 38               & NA             \\ \hline
PHY Header ($\mu$s)         & \multicolumn{3}{c|}{20}                                \\ \hline
Headers (Bytes)             & \multicolumn{3}{c|}{60}                                \\ \hline
ACK ($\mu$s)                & \multicolumn{3}{c|}{40}                                \\ \hline
$CW_{min}$                  & 15                 & 15               & NA             \\ \hline
$CW_{max}$                  & 1023               & 1023             & NA             \\ \hline
$m$                         & 6                  & 6                & NA             \\ \hline
\end{tabular}%
}
\end{table}

\subsubsection{Hyper parameters}
\begin{figure}
    \centering
    \includegraphics[width=.35\textwidth]{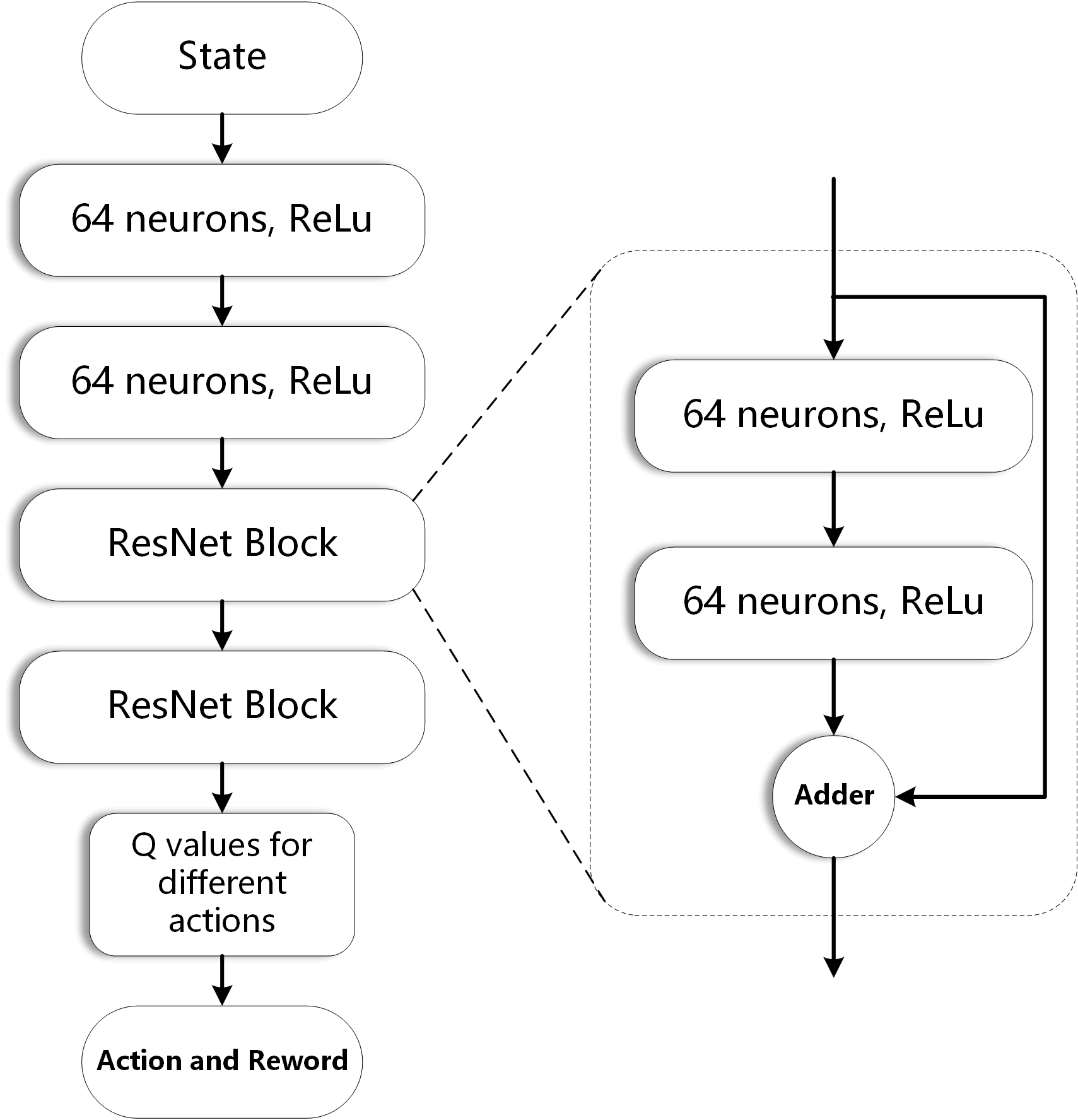}
    \caption{Deep Q-learning Neural Network. \\
    The architecture of the QNN is a six-hidden-layer ResNet, with 64 neurons in each hidden layer.}
    \label{fig:dqn}
\end{figure}
We construct a six-hidden-layer ResNet, with 64 neurons in each hidden layer for the deep Q-learning neural network illustrated in Fig \ref{fig:dqn}. Since for the system model in Section \ref{sec:sys}, the state dimension, and the input data are relatively small in the examples simulated,  we chose a six-hidden-layer ResNet with the first two fully connected layers followed by two ResNet blocks. Each ResNet block contains two fully connected hidden layers plus one "shortcut" from the input to the output of the ResNet block suggested in \cite{yu2019deep}. The network is not too deep, so that the model weight could be controlled. Table \ref{tb:hyper_pra} lists the hyperparameters of the deep Q-learning neural network.

\begin{table}[]
\caption{Hyper-Parameters of QNN}
\label{tb:hyper_pra}
\centering
\resizebox{.3\textwidth}{!}{
\begin{tabular}{|c|c|}
\hline
\textbf{Parameters}         & \textbf{Value}  \\ \hline
State size & 40 \\ \hline
Batch size & 32 \\ \hline
Learning rate & 0.001   \\ \hline
$\gamma$ in eq (\ref{eq:gam}) & 0.9 \\ \hline
$\epsilon$ in eq (\ref{eq:eps}) &  1 \\ \hline
$\epsilon$ minimum value & 0.01 \\ \hline
$\epsilon$ decay & 0.995 \\ \hline
Memory size & 1000 \\ \hline
Replace target iteration & 200 \\ \hline

\end{tabular}
}
\end{table}

\subsubsection{Pre-trained model}

Each node is assumed to have a common knowledge of the channel status - i.e., when the channel is idle, a node attempts to transmit and conversely (when busy, does not). For each simulation run, a pre-trained DRL network is used; this pre-trained model (for five stations) is based on an 80,000 step run whereupon the model converges to learn the basic knowledge of competing and cooperation.

\begin{figure}[h]
    \centering
    \includegraphics[width=.35\textwidth]{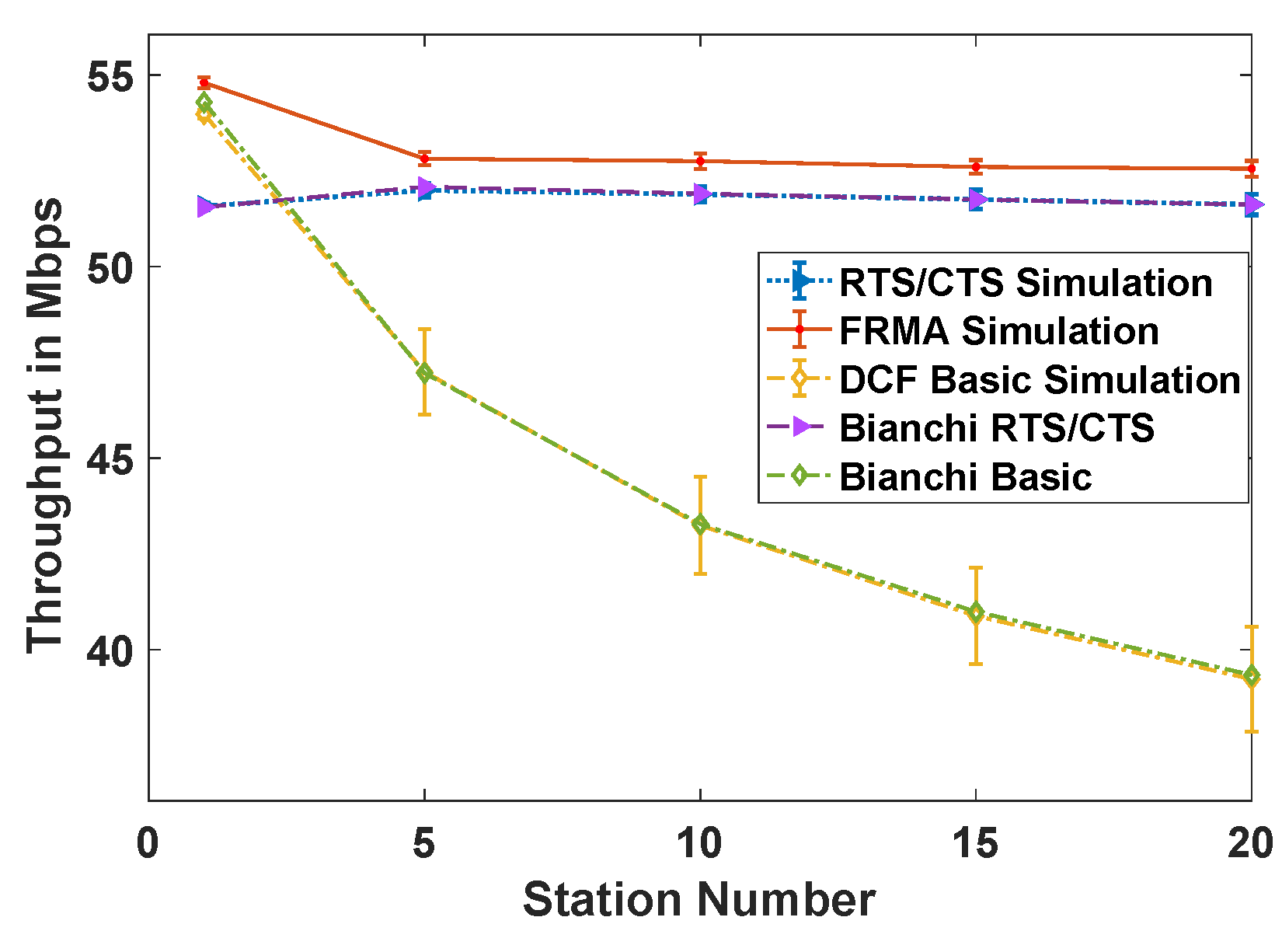}
    \caption{Average Throughput Comparison}
    \label{fig:performance}
\end{figure}

\begin{figure}
    \centering
    \includegraphics[width=.35\textwidth]{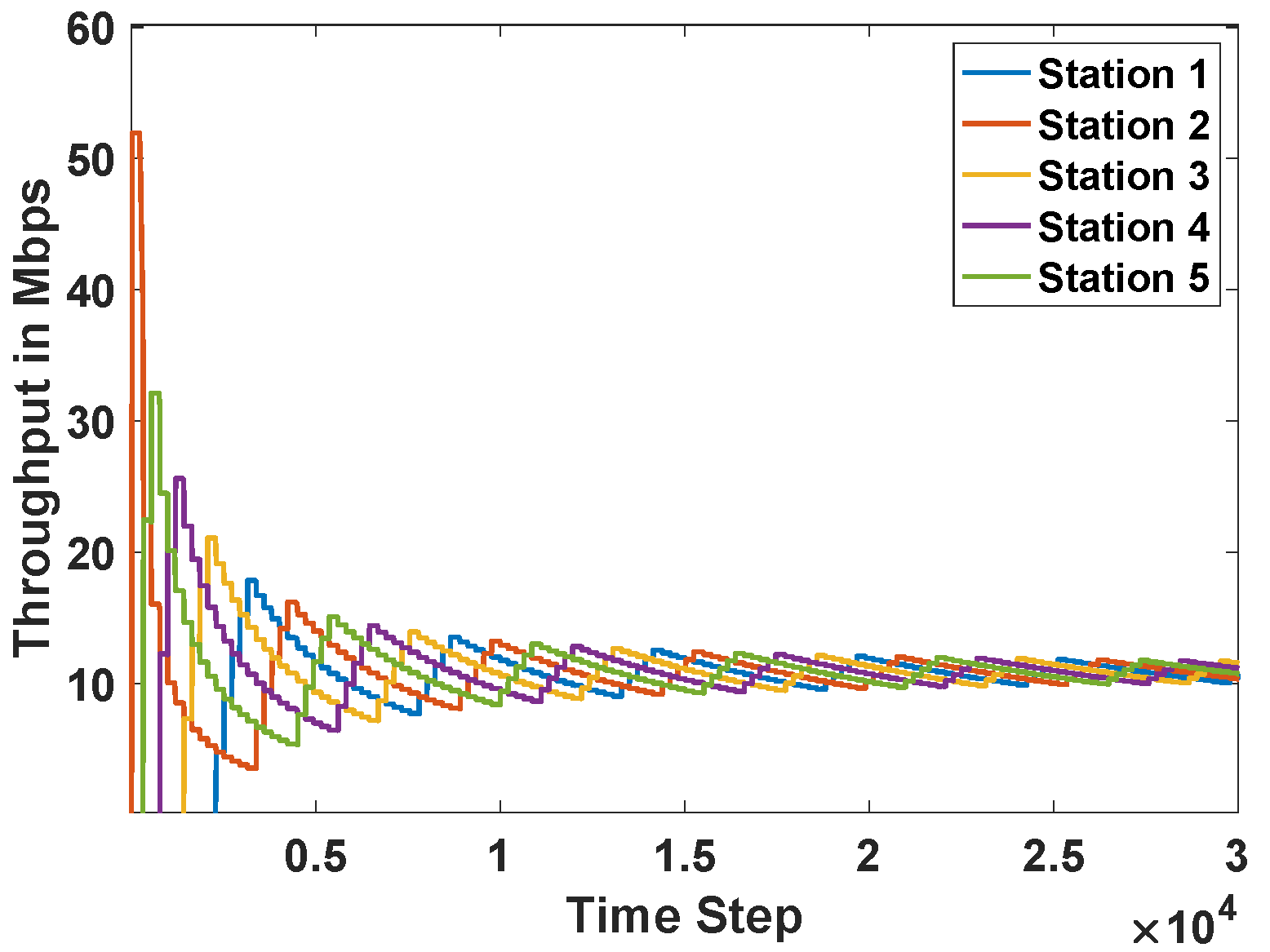}
    \caption{Individual throughput between users change with the simulation time, when station number is 5 }
    \label{fig:indivi_user}
\end{figure}

\subsection{Results}
The simulation result of throughput performance is shown in Fig. \ref{fig:performance}. For the DCF basic method and RTS/CTS method, we can see that the simulation result is very close to the theoretical Bianchi's results according to equation (\ref{eq:fraction}). Based on these validated result, we then implement the FRMA mechanism. From the simulation result, as the number of stations increases, the aggregated throughput drops for the DCF basic method because of the collision probability within each time slot increases. For the RTS/CTS mechanism, the throughput for the single user is lower than DCF basic because it has a larger overhead to wait for the RTS/CTS time while no collision happens in the single-user mode. However, as the number of stations increases, the throughput of RTS/CTS remains the same and decreases very slowly, which indicates its stability. The reason why RTS/CTS performs better than DCF basic is that only the RTS frame drops rather than an entire packet when the collision happens. In our FRMA mechanism, the training phase enables each station to be aware of the environment and stations around itself, so all stations operate collaboratively. Thus, each station evolves into a new mechanism which outperforms BEB by learning the environment. The overhead of the FRMA is the transmission of the broadcast of the federated learning. It happens once in a mean of 100 successful transmissions in this setting. Note that the overhead of FRMA is accounted for the calculation of the network throughput. The average throughput of FRMA outperforms RTS/CTS by 5\% and DCF basic by 20\%.

The simulation of fairness among users is shown in Fig. \ref{fig:indivi_user}. We use the pre-trained model in the beginning to save the training duration for practical reason. We notice that station 2 is very aggressive in the beginning, and after $10000$ time steps, the whole network converges due to the FL algorithm. Although $10000$ time steps look like a long duration, each time step takes only $10$ micro-second, so $10000$ time step is equal to $0.1$ second, which is the convergence duration. Recall that FL runs in the following manner, the AP calculates the average throughput for each user by receiving the packet and then apply the federated learning method. It will update and distribute the average weight of the neural network back to each station. Every period of $T=100$ successful transmission, the AP runs the FL algorithm until the fairness is achieved. This simulation result coincides with our previous analysis. That is, without FL, there is always one station which is the most aggressive one among all users, and the whole network will diverge to the case where only one station transmits and all other users never transmit. This phenomenon also verifies the significance of the application of the FL algorithm.

\section{Conclusion}
\label{sec:conclusion}

In this paper, we propose distributed DRL aided by FL as a multiple access solution in the WiFi network. Each station in the WiFi network is installed with a QNN and trained by its local information. On top of that, the FL algorithm on AP helps obtains the global QNN every period of time and broadcast the global QNN to all stations in the network. In the simulation section, IEEE 802.11 DCF basic and RTS/CTS are utilized as benchmarks, and their performance is lower than FRMA in terms of the network throughput. Bianchi's model is also applied to validate and calibrate the python-based simulation of DCF and RTS/CTS. At the end of the simulation section, we show the throughput of each user's throughput vs. training steps, and the proportional fairness of the network is achieved via the FL algorithm.

\bibliographystyle{IEEEtran}
\bibliography{reference}
\end{document}